\documentclass[a4paper,11pt]{article}
\pdfoutput=1
\usepackage{jheppub}

\usepackage{hyperref}
\hypersetup{
     colorlinks=true,       
     linkcolor=red,          
     citecolor=magenta,        
     filecolor=magenta,      
     urlcolor=blue           
}

\usepackage{graphicx}
\usepackage{placeins}
\usepackage{url}
\usepackage{multirow}
\usepackage{paralist}
\usepackage{amssymb}
\usepackage{amsmath}
\usepackage{xspace}
\usepackage{booktabs}
\usepackage[symbol]{footmisc}

\usepackage{lineno}

\newcommand{\newc}{\newcommand}
\newc{\cpp}{\textsf{C++}}
\newc{\HWS}{\textsf{Herwig 7}}
\newc{\HW}{\textsf{Herwig}}
\newc{\HSeven}{\textsf{H7}}
\newc{\HWHML}{\textsf{H7+HADML}}
\newc{\ThePEG}{\textsf{ThePEG}}
\newc{\HADML}{\textsf{HADML}}
\newc{\HWPPClass}[1]{\mbox{\href{http://projects.hepforge.org/herwig/doxygen/classHerwig_1_1#1.html}{\textsf{#1}}}}
\newc{\NC}{N_{\mathrm{c}}}

\begin{document}
\newcommand{\xju}[1]{\textcolor{red}{#1}}
\newcommand{\bpn}[1]{\textcolor{red}{#1  --bpn}}

\title{Fitting a Deep Generative Hadronization Model}

\affiliation[a]{Department of Physics, University of Wisconsin-Madison, Madison, WI 53706, USA}
\affiliation[b]{Physics Division, Lawrence Berkeley National Laboratory, Berkeley, CA 94720, USA}
\affiliation[c]{Berkeley Institute for Data Science, University of California, Berkeley, CA 94720, USA}
\affiliation[d]{Department of Physics, University of California, Berkeley, CA 94720, USA}
\affiliation[e]{Jagiellonian University, Krakow, Poland}
\author[a,b]{Jay Chan,}
\author[b]{Xiangyang Ju,}
\author[e]{Adam Kania,}
\author[b,c]{Benjamin Nachman,}
\author[d,b]{Vishnu Sangli,}
\author[e]{and Andrzej Siodmok}

\abstract{
Hadronization is a critical step in the simulation of high-energy particle and nuclear physics experiments.  As there is no first principles understanding of this process, physically-inspired hadronization models have a large number of parameters that are fit to data.  Deep generative models are a natural replacement for classical techniques, since they are more flexible and may be able to improve the overall precision.  Proof of principle studies have shown how to use neural networks to emulate specific hadronization when trained using the inputs and outputs of classical methods.  However, these approaches will not work with data, where we do not have a matching between observed hadrons and partons.  In this paper, we develop a protocol for fitting a deep generative hadronization model in a realistic setting, where we only have access to a set of hadrons in data.  Our approach uses a variation of a Generative Adversarial Network with a permutation invariant discriminator.  We find that this setup is able to match the hadronization model in \textsc{Herwig} with multiple sets of parameters.  This work represents a significant step forward in a longer term program to develop, train, and integrate machine learning-based hadronization models into parton shower Monte Carlo programs.
}

\maketitle

\section{Introduction}
\label{sec:intro}

Hadronization connects theory and experiment by transforming the fundamental degrees of freedom -- quarks and gluons -- with observable degrees of freedom -- hadrons.  However, we do no have a first-principles understanding of hadronization and so existing approaches use physically-inspired, highly flexible models fit to data.  Our vision is to replace these hand-crafted models with deep learning, where the additional exprresivity would have the potential to enhance precision, the models would be readily differentiable, and they would be naturally compatible with Graphical Processing Unit (GPUs).

There are currently two hadronization models in wide use: the cluster model~\cite{Webber:1983if} and the string model~\cite{Andersson:1983ia,Sjostrand:1984ic}.  The former is employed by default in the Herwig~\cite{Corcella:2000bw,Bahr:2008pv,Bellm:2015jjp,Bellm:2019zci} and Sherpa~\cite{Gleisberg:2008ta,Sherpa:2019gpd} Parton Shower Monte Carlo (PSMC) programs and the latter is used by default in the Pythia~\cite{Sjostrand:2007gs,Sjostrand:2006za} PSMC.  Previously, Refs.~\cite{Ilten:2022jfm} and~\cite{Ghosh:2022zdz} showed that deep generative models could emulate the string and cluster models, respectively, in a simple setting where the neural network has access to parton-hadron pairs and only pions are produced~\footnote{Hadron type was taken from Herwig.}.  Furthermore, these models were integrated into the Pythia and Herwig PSMC programs.  These papers marked an important milestone, but represent only the first steps along a multiyear program to achieve a complete, integrated, and tuned machine learning (ML)-based hadronization model.

While previous work has shown that neural networks can emulate the existing hadronization models, we want to eventually fit the models to data.  A fundamental challenge with using data directly is that hadronization acts locally on partons while only non-local information about hadrons is observable.  In other words, events are measured as a permutation-invariant set of hadrons that have no inherent order or grouping to know which hadrons `came from' the same partons.  This means that we need a model that can learn to generate hadrons from partons based on information from a loss function that acts on the set of observable hadrons.  

The two-level challenge of fitting to data rules out most standard implementations of deep generative models.  Variational Autoencoders (VAE)~\cite{kingma2014autoencoding,Kingma2019}, Normalizing flows (NF)~\cite{10.5555/3045118.3045281,Kobyzev2020}, and diffusion models~\cite{diff1, diff2, diffvsgan} do not directly apply because we need to know the probability density of the partons and we need a permutation invariant reconstruction loss (VAE), probability density (NF), or score function (diffusion).  While there has been some progress on these fronts~\cite{Howard:2021pos,klein2022flows,Mastandrea:2022vas,Kansal:2021cqp,Buhmann:2023pmh,Kach:2022qnf,Verheyen:2022tov,Leigh:2023toe,Mikuni:2023dvk,Buhmann:2023bwk}, Generative Adversarial Networks (GANs)~\cite{Goodfellow:2014:GAN:2969033.2969125,Creswell2018} can be naturally applied to this setting.  For GANs, the latent space does not require a tractable probability density, the discriminator can be applied on a different level (hadrons) as the generator (partons), and permutation invariance can be enforced by using a set-based classifier for the discriminator.  GANs were the first deep generative model applied to particle physics data~\cite{deOliveira:2017pjk,Paganini:2017dwg,Paganini:2017hrr} and have since been extensively studied (see e.g. Ref.~\cite{Adelmann:2022ozp,Butter:2022rso,hepmllivingreview}).  GAN-like setups have also been used for two-level fitting in the context of parameter estimation~\cite{Andreassen:2020gtw} and unfolding~\cite{momentunfolding}.  We propose to use GANs for fitting hadronization models to data.

We embed the GAN-based hadronization model HadML introduced in Ref.~\cite{Ghosh:2022zdz} in a full event-level fitting framework.  A fully connected neural network takes as input individual clusters and outputs pairs of hadrons.  This network acts in the cluster rest frame.  The resulting hadrons are then boosted to the lab frame and the GAN discriminator is based on Deep Sets~\cite{zaheer2018deep}, which is a permutation invariant neural network architecture.  We restrict ourselves to the cluster model inputs (clusters created from pre-confined partons) and pion outputs in order to focus on the two-level fitting challenge.  These simplifications will be relaxed in future work.

This paper is organized as follows.  Section~\ref{sec:methods} introduces the conceptual and technical details behind our fitting framework.  Numerical examples are presented in Sec.~\ref{sec:results}, including two variations on the cluster model.  The paper ends with conclusions and outlook in Sec.~\ref{sec:conclusion}.

\section{Methods}
\label{sec:methods}

\subsection{Statistical Approach}

Our goal is to learn a conditional generator function $G\left(z, \lambda; \omega_G \right)$ which maps cluster kinematic properties onto the kinematic properites of the two\footnote{The cluster model can produce more than two hadrons, but most of the time, at the energies we consider, there are only two.  We restrict to two for this study and will explore more complex decays in future work.} hadrons from each cluster decay $\{h_1, h_2\} \in\mathbb{R}^{2 N_h}$ with the parameters $\omega_G$. Here, $z\in\mathbb{R}^{N_z}$ is the input noise variable sampled from the prior $p\left(z\right)$, and $\lambda\in\mathbb{R}^{N_\lambda}$ is the conditional variable, namely the cluster kinematic properties. Since two hadrons from a cluster decay must be back-to-back in the rest frame of cluster, the generator $G$ can instead output the polar angles $\theta$ and $\phi$ of the ``first hadron'' in the cluster rest frame. Note that here $\phi$ is defined in the range of $(-\pi/2,\ \pi/2)$, and the hadron with $\phi$ in this range is defined to be the first hadron. In the original setup \cite{Ghosh:2022zdz}, a discriminator function $D\left(\theta, \phi; \omega_D \right)$, parametrized with $\omega_D$, is learned to represent the probability that $\{\theta, \phi\}$ came from cluster fragmentation rather than the generator $G$. $G$ and $D$ are then trained alternately to maximize and minize the loss function, respectively:

\begin{equation}
    L = - \sum_{\lambda \sim \textsc{Herwig},\, z \sim p\left(z\right)} \left( \log\left(D\left(\tau\left(\lambda\right)\right)\right) + \log\left(1 - D\left(G\left(z, \lambda\right)\right)\right) \right)\,,
\end{equation}
where $\tau$ is the cluster fragmentation.

\begin{figure}[h!]
    \centering
    \includegraphics[width=0.95\textwidth]{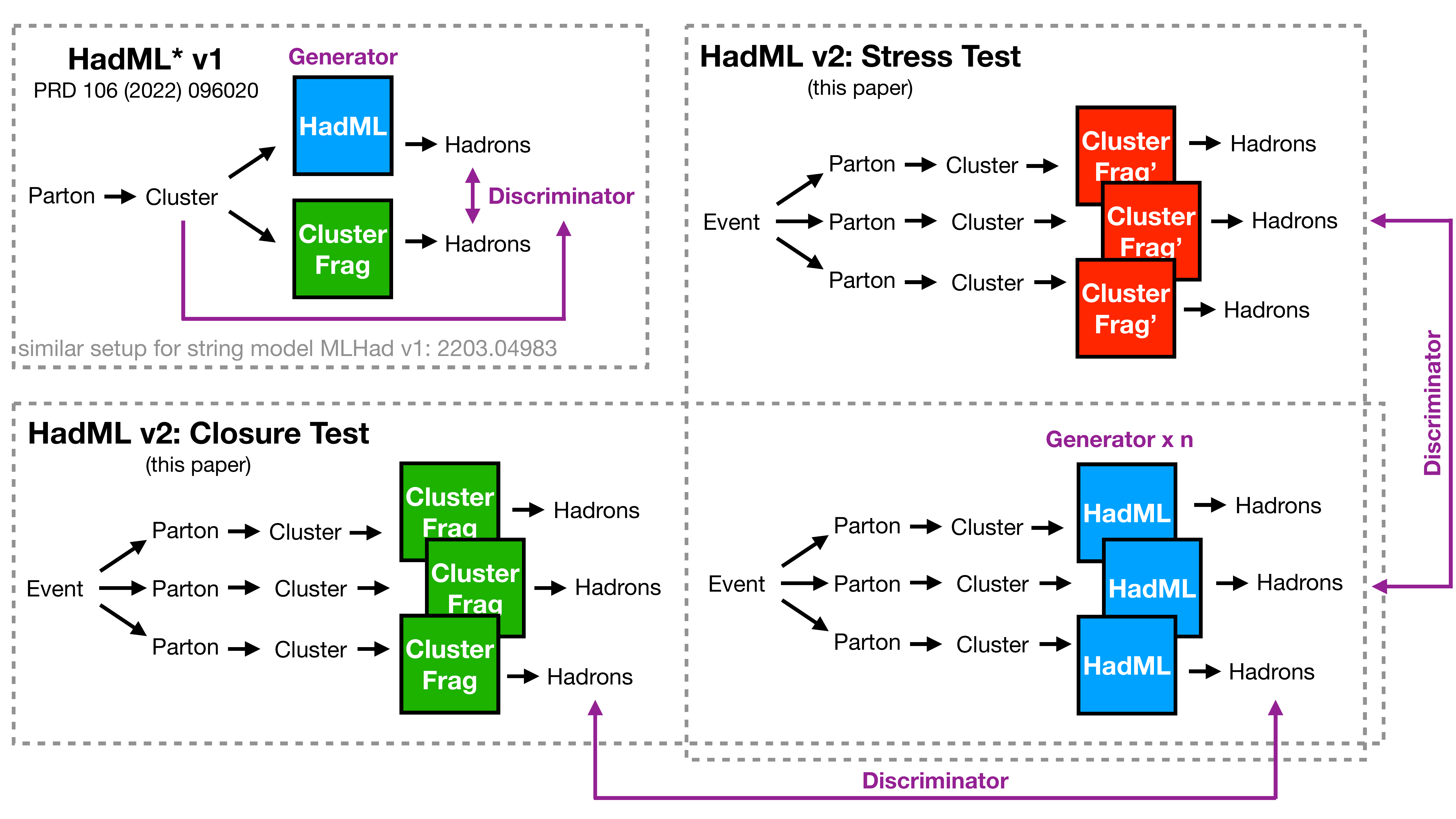}
    \caption{An overview of the model presented in this paper and how it compares to HadML v1 from Ref.~\cite{Ghosh:2022zdz}.  Since the clusters are not observable in data, the discriminator in v2 acts on sets of hadrons and does not have access to cluster-hadron-hadron labels.  We first study the performance in the same Herwig setup as in Ref.~\cite{Ghosh:2022zdz} (`Closure Test') and then check that it is also able to fit another Herwig setup (Cluster Frag') with variations in the cluster hadronization model (`Stress Test').}
    \label{fig:schematic}
\end{figure}

In the setup above, all hadrons are paired and matched to a cluster. In the actual data, however, the only observables are the kinematic properties of each individual hadron. In order to be able to fit the model to actual data, where the hadron matching and cluster information is not accessible, the discriminator function is modified to be $D_E\left(x \right)$, where $D_E$ takes a set of hadron kinematic properties $x \equiv \{h_1, h_2,..., h_n\}$ in the same event as inputs. Furthermore, we parameterize $D_{E}$ as a Deep Sets model~\cite{zaheer2018deep}:
\begin{equation}
    D_E\left(x \right) = F\left(\frac{1}{n}\sum_{i=1}^n \Phi\left(h_i, \omega_{D_\Phi}\right), \omega_{F}\right)\,,
\end{equation}
where $\Phi$ embeds a set of hadrons into a fixed-length latent space and
$F$ acts on the average of the latent space. Due to the average, $D_E$ can take any length of hadron set and is invariant under permutations of hadrons. The loss function thus becomes:

\begin{equation}
    L = - \sum_{x \sim \text{data}} \log\left(D_E\left(x\right)\right) - \sum_{\left\{G\right\} \sim \textsc{Herwig},\, z \sim p\left(z\right)} \log\left(1 - D_E\left(\left\{G\left(z, \lambda\right)\right\}\right)\right)\,,
\end{equation}
where $\left\{G\left(z, \lambda\right)\right\}$ is generated by a set of clusters that came from the same event.  The generator acts in the cluster rest frame and then the resulting hadrons are boosted into the lab frame before being passed to the discriminator.  A summary of the setup and how it differs from Ref.\cite{Ghosh:2022zdz} is presented in Fig.~\ref{fig:schematic}.

In our implementation, $G$ is a neural network.  However, this approach could also be used to fit (without binning) data to a parametric physics model as well.  For that case, $G$ would be e.g. the cluster model and the parameters would not be weights and biases of a neural network, but instead the parameters of the cluster model.  This would require making the cluster model differentiable so that gradients could be passed through the model.  We leave explorations of this hybrid setup to future work.

\subsection{Machine Learning Implementation}

Both the generator and discriminator functions are parametrized as neural networks and implemented using PyTorch \cite{NEURIPS2019_9015}. The generator is a fully connected network which consists of two hidden layers with 256 nodes per layer. The noise dimension is set to 10. The discriminator comprises two networks $\phi$ and $F$. Both $\phi$ and $F$ are a fully connected network with two hidden layers of 256 nodes each. Each intermediate layer in these networks uses a batch normalization and a LeakyReLU~\cite{xu2015empirical} activation function. The last layer of the generator uses a $\tanh$ activation function to restrict the outputs to be in the range of $(-1,\ 1)$. The outputs are then scaled and transformed linearly to match the actual range $(-\pi/2,\ \pi/2)$ for $\phi$ and $(0,\ \pi)$ for $\theta$. The last layer of $F$ uses a sigmoid activation function and no activation is used for the last layer of $\Phi$.  

All neural network inputs are normalized to the range of $(-1,\ 1)$, whereas the noise prior $p$ is a Gaussian distribution with a mean of 0 and width of 1. The generator and discriminator are optimized alternately (1 discriminator step and 5 generator steps) with Adam \cite{adam} with a learning rate of $5 \times 10^{-7}$ and $10^{-4}$ for the generator and discriminator, respectively. The training uses a batch size of 10,000 and is performed for 6,000 epochs.  The hyperparameters were optimized with Weights and Biases~\cite{wandb}.

\section{Results}
\label{sec:results}

\subsection{Datasets}
\label{sec:dataset}
Crucial data for fitting hadronisation models are LEP events collected in 
$e^+e^-$ collisions at the center-of-mass energy $\sqrt{s}=91.2$ GeV.
Therefore, we used such events generated with version 7.2.1 of the Herwig Monte Carlo
generator for a training dataset for our Generative Hadronization Model. As mentioned earlier, the cluster 
model~\cite{Webber:1983if} is used for hadronisation in the Herwig
generator. Based on the color preconfinement~\cite{Amati:1979fg}, the cluster model 
groups a partonic final state into a set of colour-singlet clusters (pre-hadrons) with an invariant mass distribution that is independent of the specific hard scattering process or its centre-of-mass energy and that peaks at low masses. Therefore, most clusters decay into two hadrons. However, a small fraction of clusters are too heavy 
for this approach to be justified. Therefore, these heavy clusters are first split into lighter clusters before decaying. The decay of such massive clusters is not discussed in this publication but will be considered in future work. Each entry in our training data set includes information about the four-momentum of all the light clusters in an event and the four-momenta of their parents (partons) and children (hadrons), along with their flavours. An example of an entry from our data sets is available on Zenodo at Ref.~\cite{jay_chan_2023_7958362}.
To simplify the training data further, only decays into $\pi$ mesons were considered\footnote{In Herwig, this is achieved by adding the following line: \textsf{set HadronSelector:Trial 1}
into the default LEP.in input card. The only other modification to 
the default hadronisation settings was the change that the hadrons 
produced from cluster decays were on the mass shell. This can be achieved by adding the command: \textsf{set ClusterDecayer:OnShell Yes} in the input file.}. 
To check whether the model can adapt to different variants of the kinematics of hadron decays, 
we also prepared two datasets with different, minimal ($0$) and maximal ($2$) settings of the {\bf ClSmr} parameter. The {\bf ClSmr} parameter is the main parameter governing the kinematics of cluster hadron decay.
 Hadrons that contain a parton produced in the perturbative stage of the event retain 
 the direction of the parton in the cluster rest frame with possible Gaussian smearing of the direction.
  The smearing is controlled by the the {\bf ClSmr} parameter through an angle $\theta_{\rm smear}$
  where
\begin{equation}
\cos\theta_{\rm smear} = 1+ {\bf ClSmr}\log\mathcal{R}.
\label{eqn:hadronsmear}
\end{equation}
where $\mathcal{R}$ is a uniform random number chosen from $[0, 1]$. For more details about the parameters
of the cluster model implemented in Herwig, see Chapter 7 of the generator's manual~\cite{Bahr:2008pv}.

In Sec.~\ref{sec:fits} we use the minimal {\bf ClSmr} as our alternative sample and refer to this setup as Herwig Cluster $kin^{min}$.  As would be the case with actual data, we use clusters from the nominal setting when fitting the alternative sample, although changing {\bf ClSmr} does not change the cluster kinematic properties and thus the inputs to the GAN model are statistically correct.  When we fit the nominal sample, the cluster inputs to the fit are distinct but statistically identical to those in the dataset we are fitting.

\subsection{Fitted Models}
\label{sec:fits}

\begin{figure}
    \centering
    \includegraphics[width=0.6\textwidth]{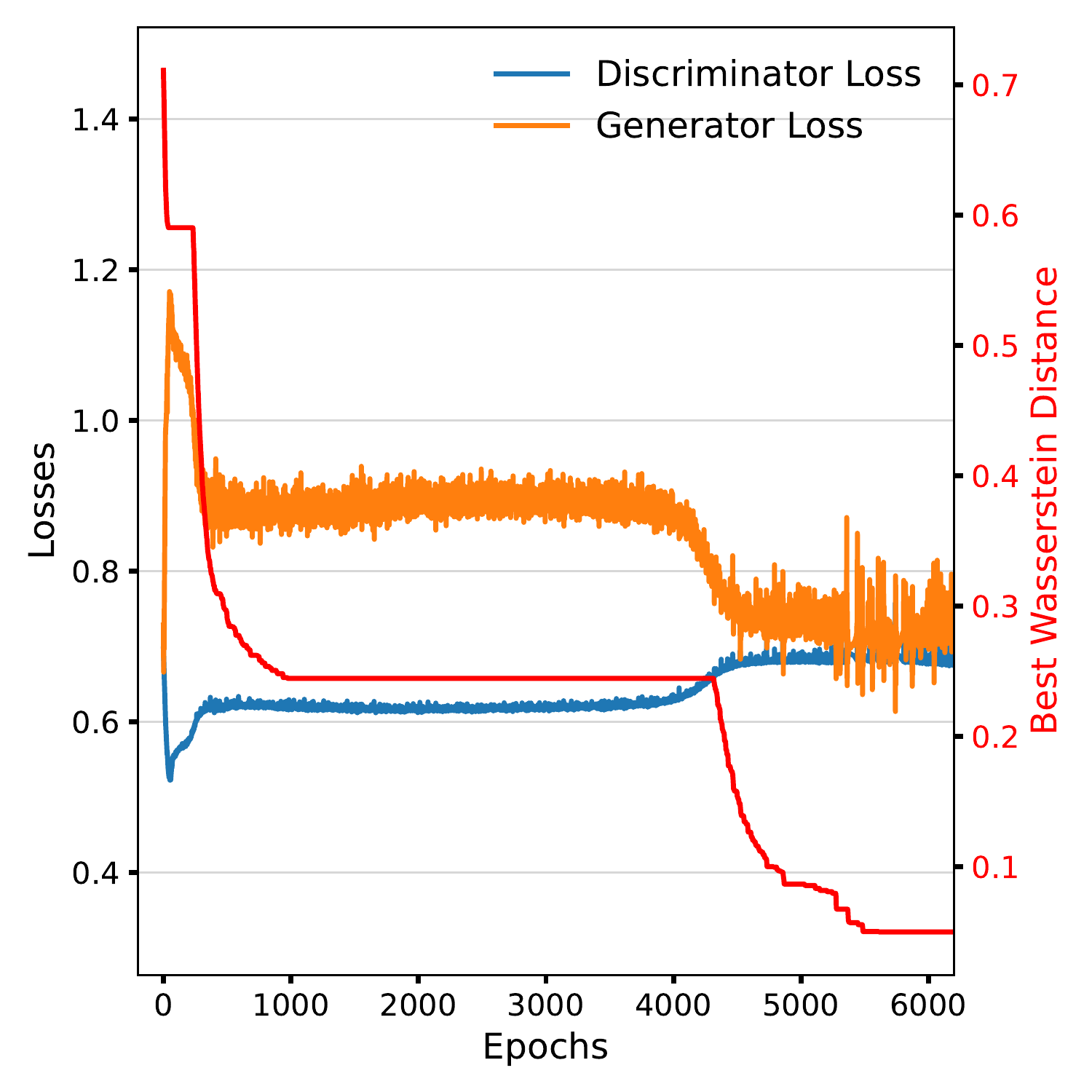}\\
    \caption{Generator loss, discriminator loss and running best Wasserstein distance as a function of the training epoch. The running best Wasserstein distance is quantified by the $y$ axis on the right side of the plot.}
    \label{fig:training}
\end{figure}

The training history of the fit is presented in Fig.~\ref{fig:training}.  As expected, the discriminator loss increases and the generator loss decreases, with a final value near $\log(2)$ (classifier outputs 0.5 for all examples).  As an independent evaluation of the model performance, we also compute the Wasserstein distance between the true and generated four-momenta in the lab frame that are used by the discriminator to update the generator. The Wasserstein distance is computed as the average over the first Wasserstein distance for each four-vector component with Scipy \cite{2020SciPy-NMeth}.  Interestingly, the best Wasserstein distance decreases for the first 1000 epochs, then plateaus for the next 3000 epochs, before dropping to the final value around 5500 epochs. There are many possible variations on the GAN training setup that are possible to further improve the performance and we plan to explore these in the future.

\begin{figure}
    \centering
    \includegraphics[width=0.5\textwidth]{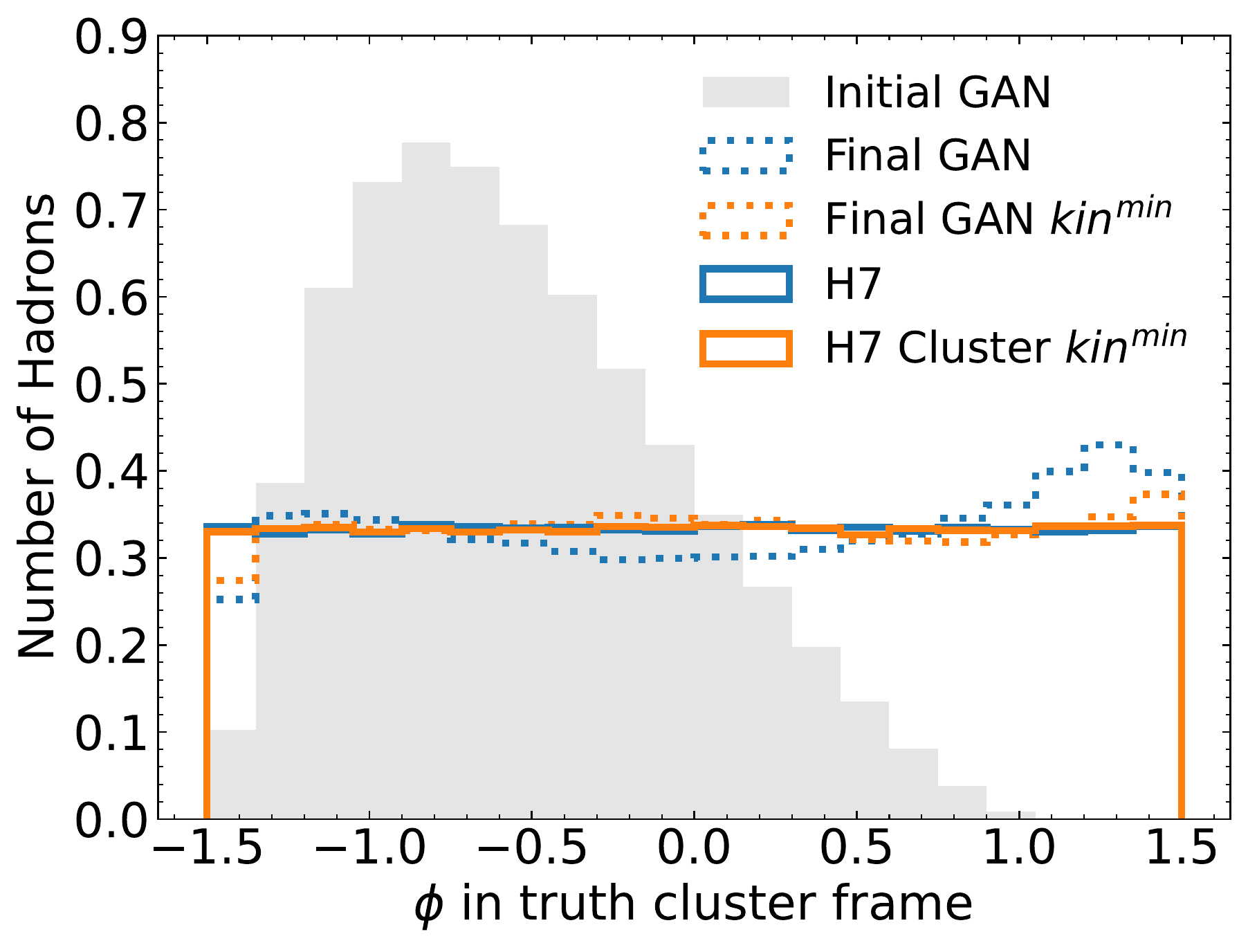}\includegraphics[width=0.5\textwidth]{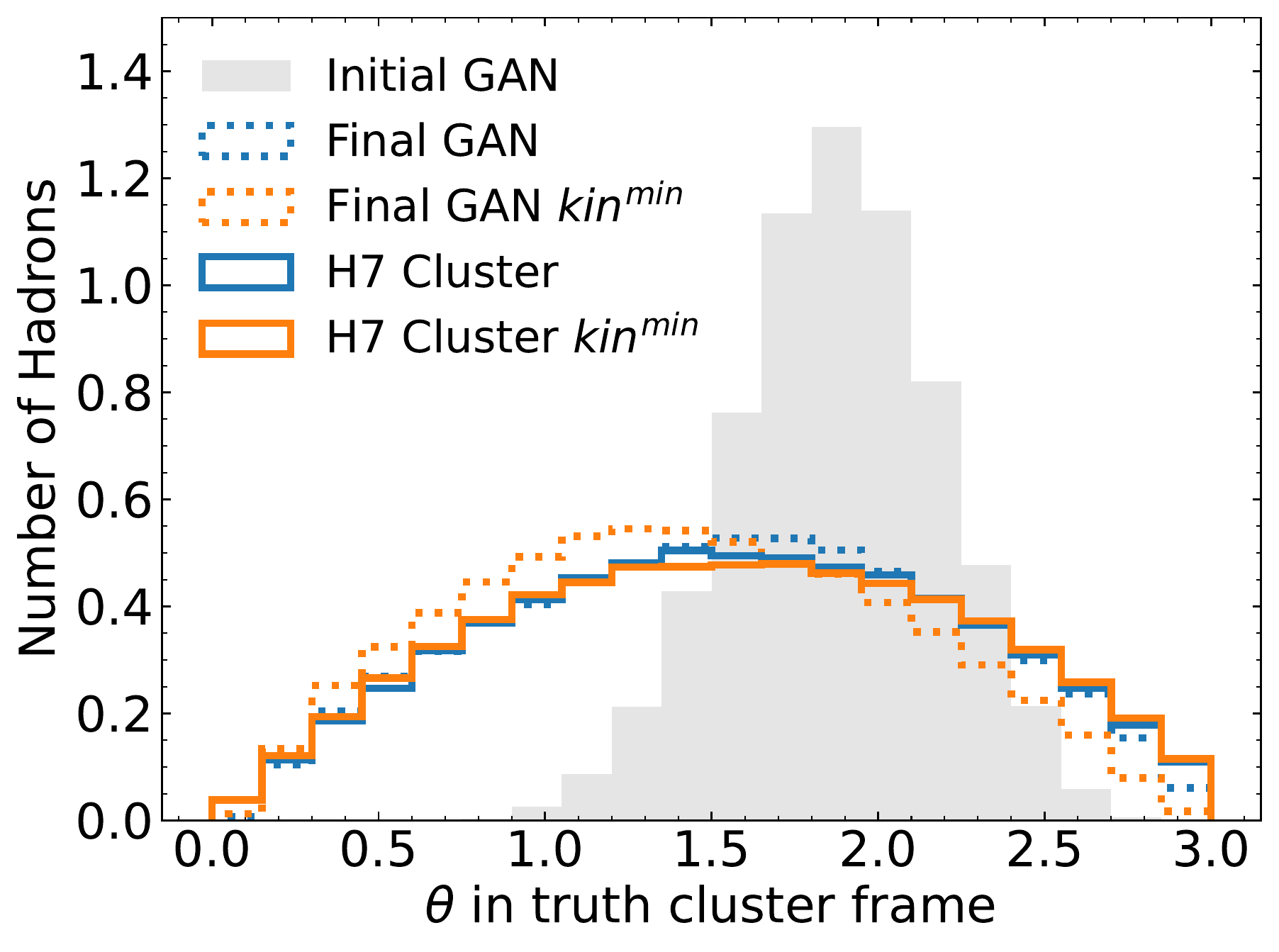}
    \\\includegraphics[width=0.5\textwidth]{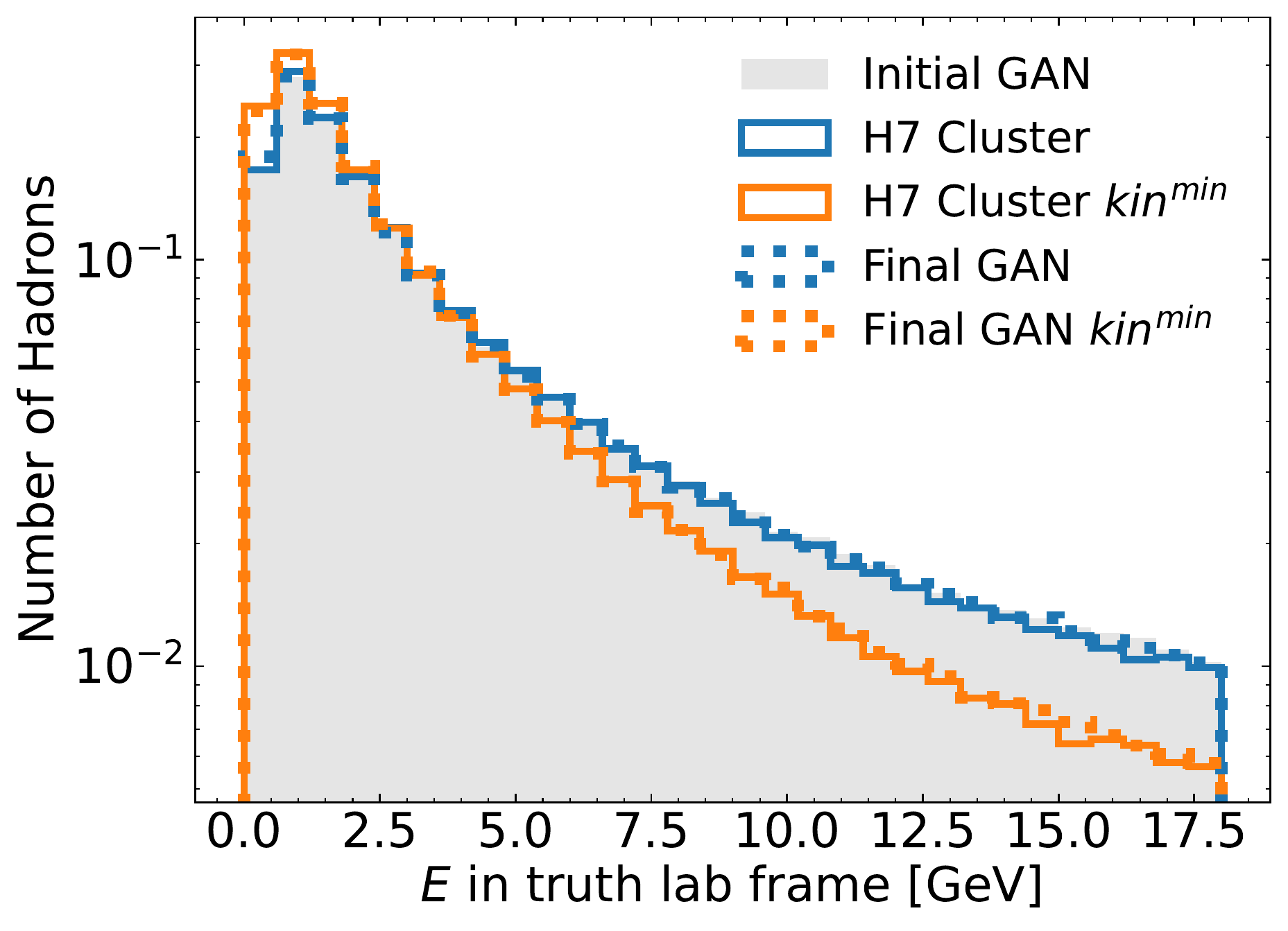}\includegraphics[width=0.5\textwidth]{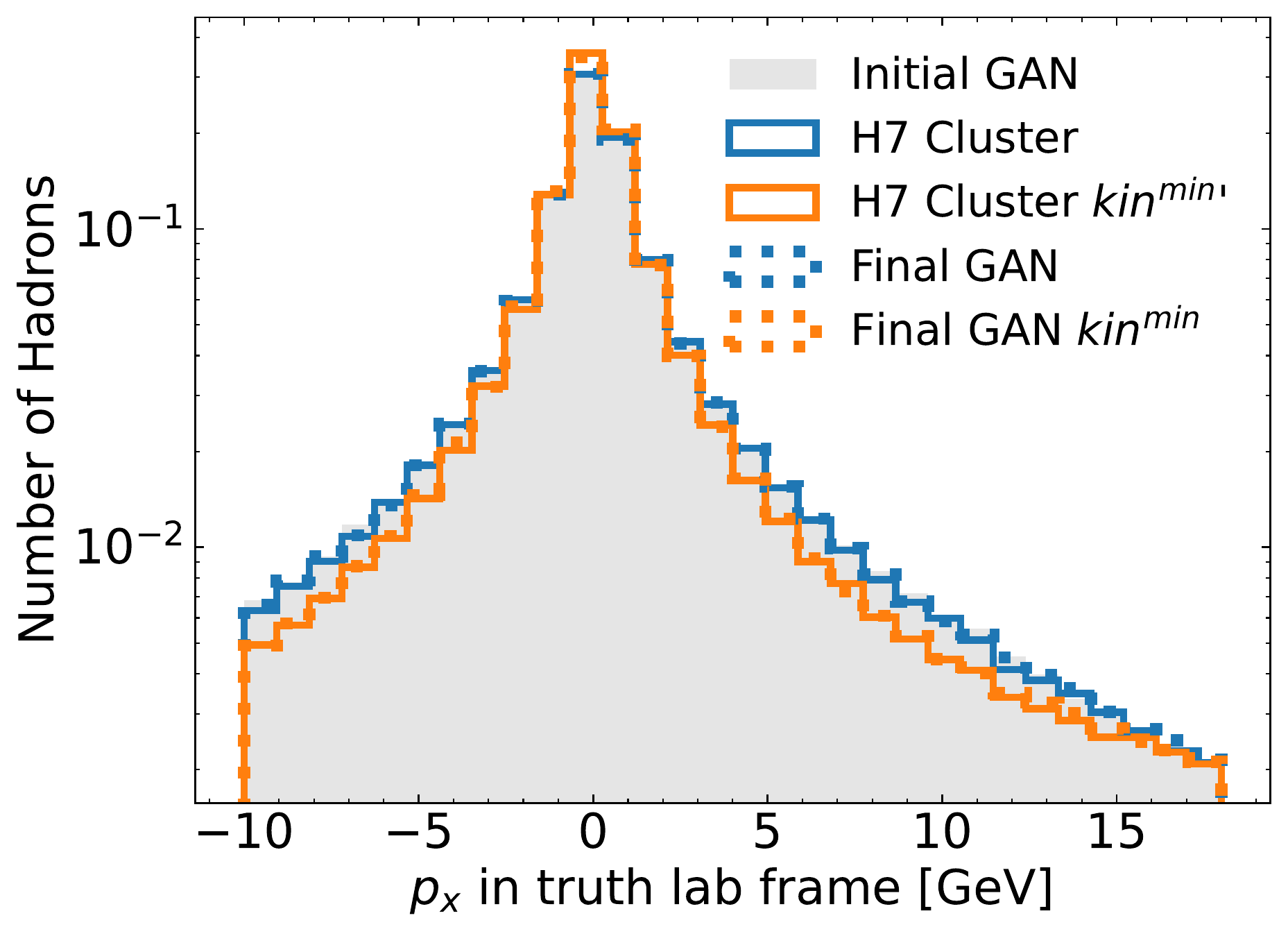}\\
    \caption{Top: the generative model in the true cluster rest frame.  Bottom: two of the four-vector components that are used by the discriminator to update the generator.}
    \label{fig:fits_inputs}
\end{figure}

The direct inputs and outputs of the model are shown in Fig.~\ref{fig:fits_inputs}.  The generator produces two outputs per cluster, corresponding to the angle of one of the pions in the cluster rest frame in spherical coordinates.  Histograms corresponding to this model are shown in the top row of Fig.~\ref{fig:fits_inputs}.  The marginal distributions looks similar to isotropic decays.  For illustration, we also show what an initialized, untrained GAN looks like in both coordinates.  The fact that the initial GAN is so far from the final GAN is a non-trivial demonstration of the learning.  Both GAN models match their respective truth Herwig spectra well.  The marginal $\phi$ distribution is uniform, which is difficult for generative models to reproduce exactly.  In the future, it may be possible to make this more precise by constructing the model to give a uniform marginal.

After the clusters are decayed, the resulting hadron kinematic properties are Lorentz boosted to the lab frame and then aggregated over all clusters in the event.  The second row of Fig.~\ref{fig:fits_inputs} shows histograms of the resulting hadron four-vectors, which are the inputs to the discriminator.  We only show the energy $E$ and the $x$ momentum $p_x$, but similar trends hold for $p_y$ and $p_z$.  Since hadronization is a small correction for such inclusive observables, the kinematic properties are mostly set by the Herwig parton shower, which is the same for the Herwig and GAN lines in the plots (since the GAN takes the clusters from the parton shower as input).  This is the reason why the initial GAN starts so close to Herwig truth.  However, the alternative Herwig sample differs significantly from the nominal Herwig sample, in particular in how hadrons split energy, which is most clearly seen in the tails of the energy and momentum distributions.  The GAN model is an excellent match to the Herwig events across the full spectra.

\begin{figure}
    \centering
    \includegraphics[width=0.5\textwidth]{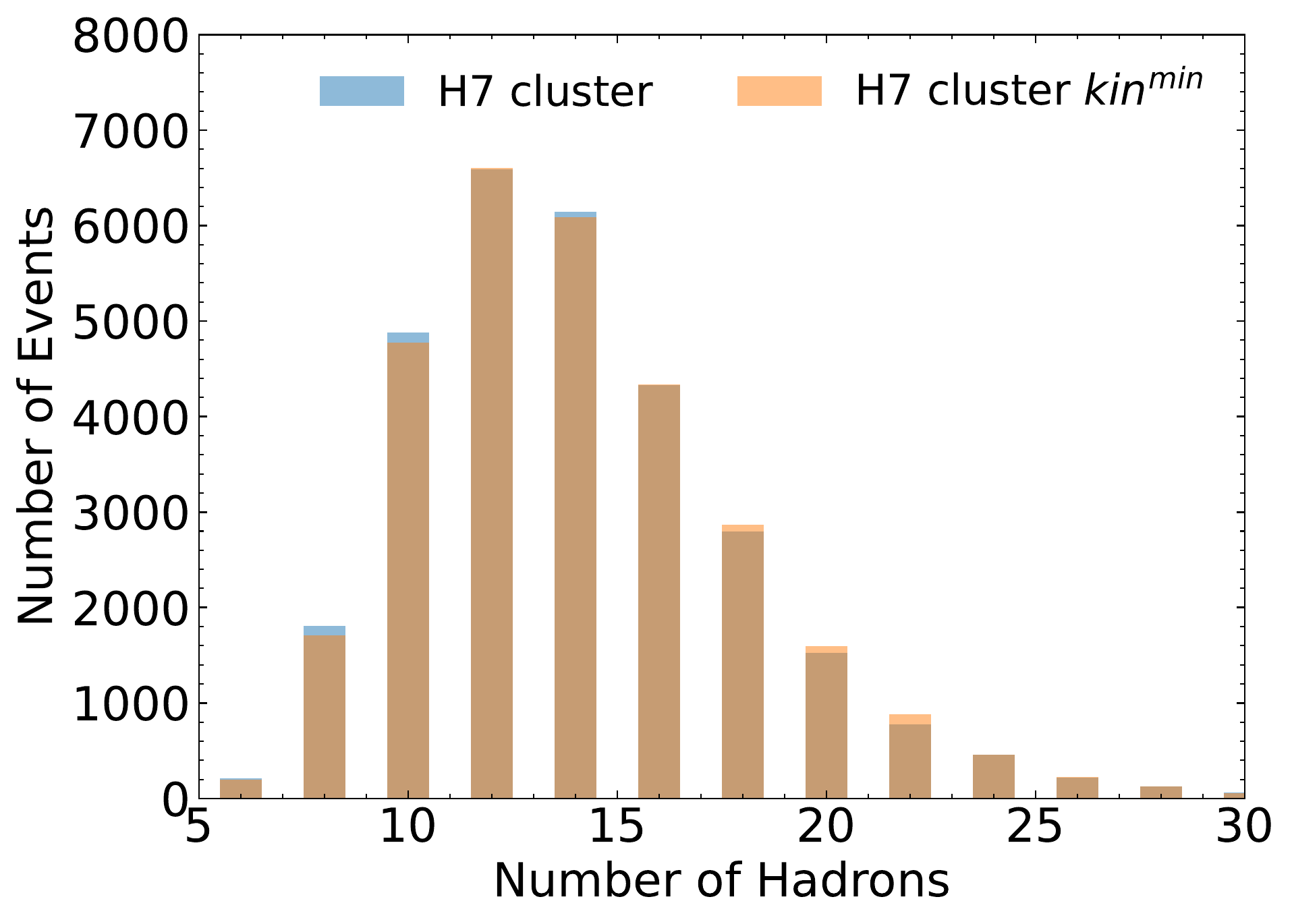}\\
    \includegraphics[width=0.5\textwidth]{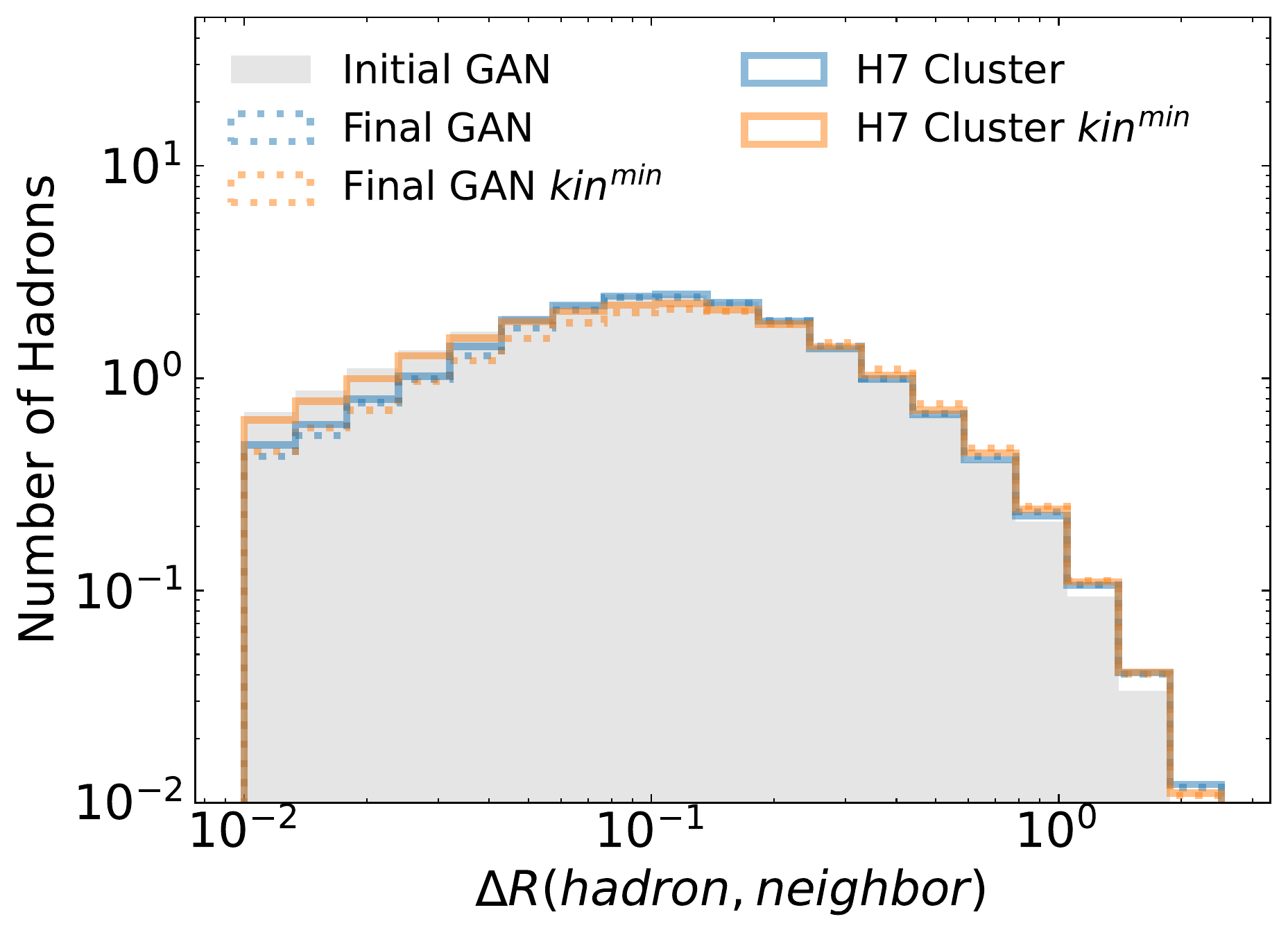}\includegraphics[width=0.5\textwidth]{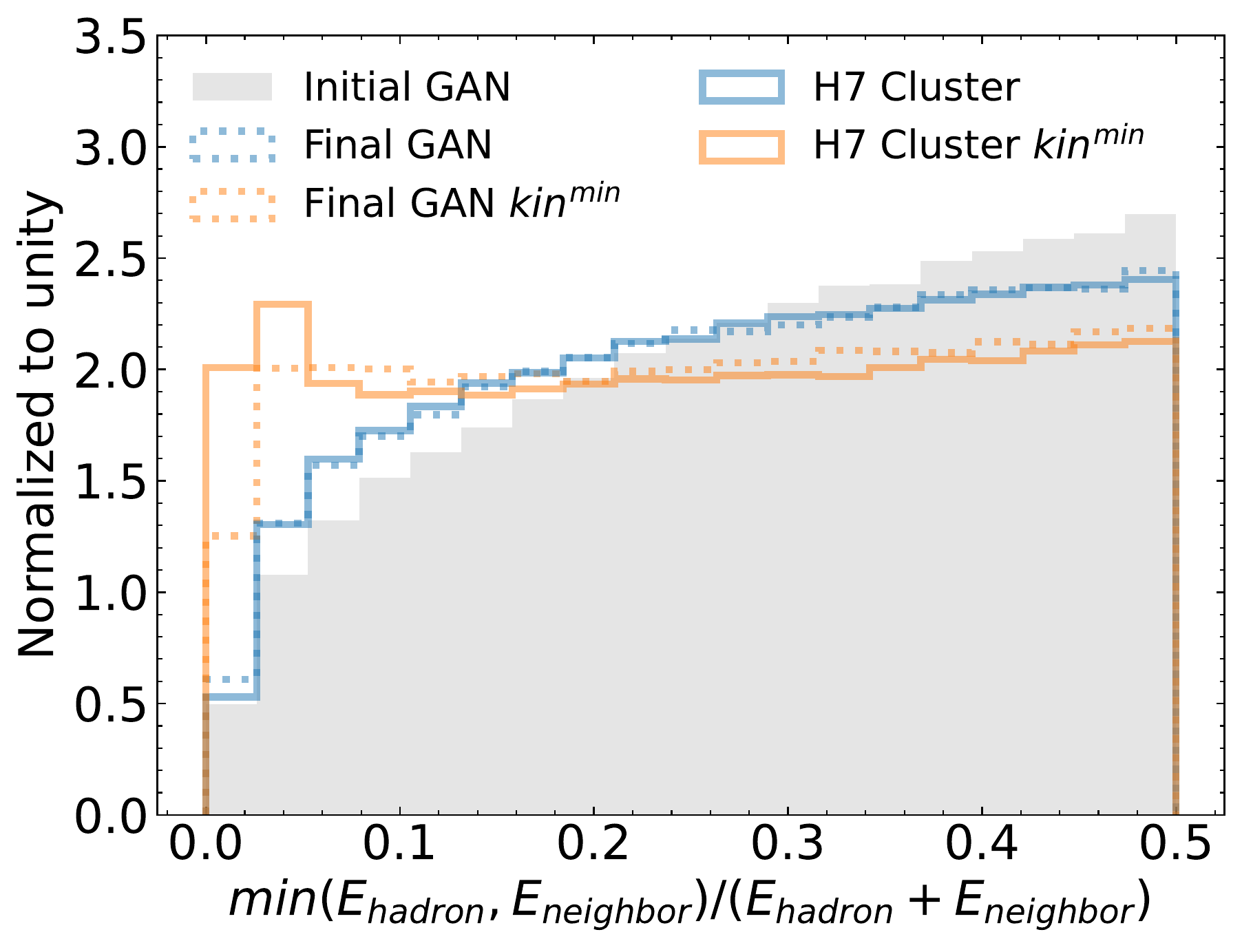}\\
    \includegraphics[width=0.5\textwidth]{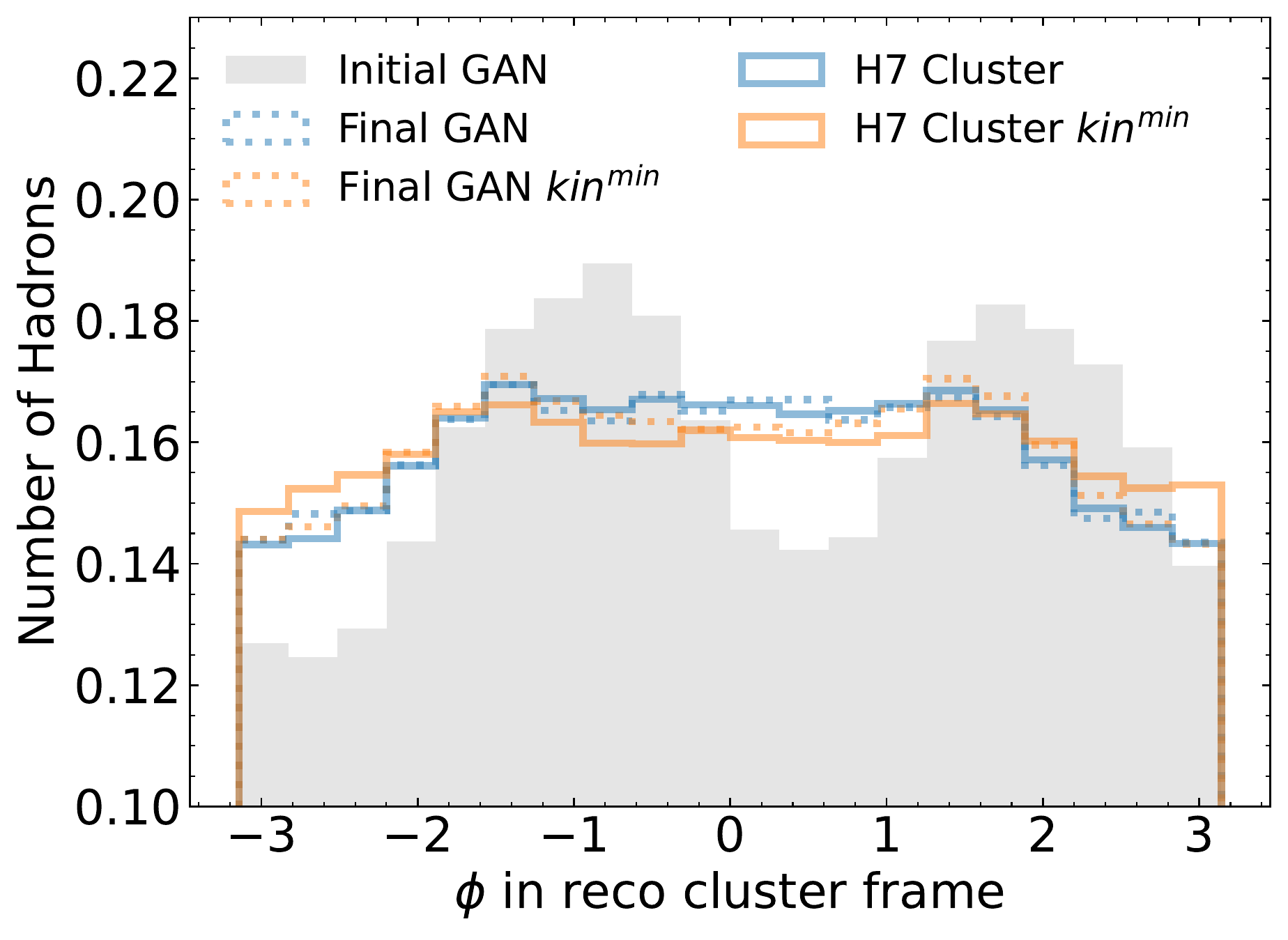}\includegraphics[width=0.5\textwidth]{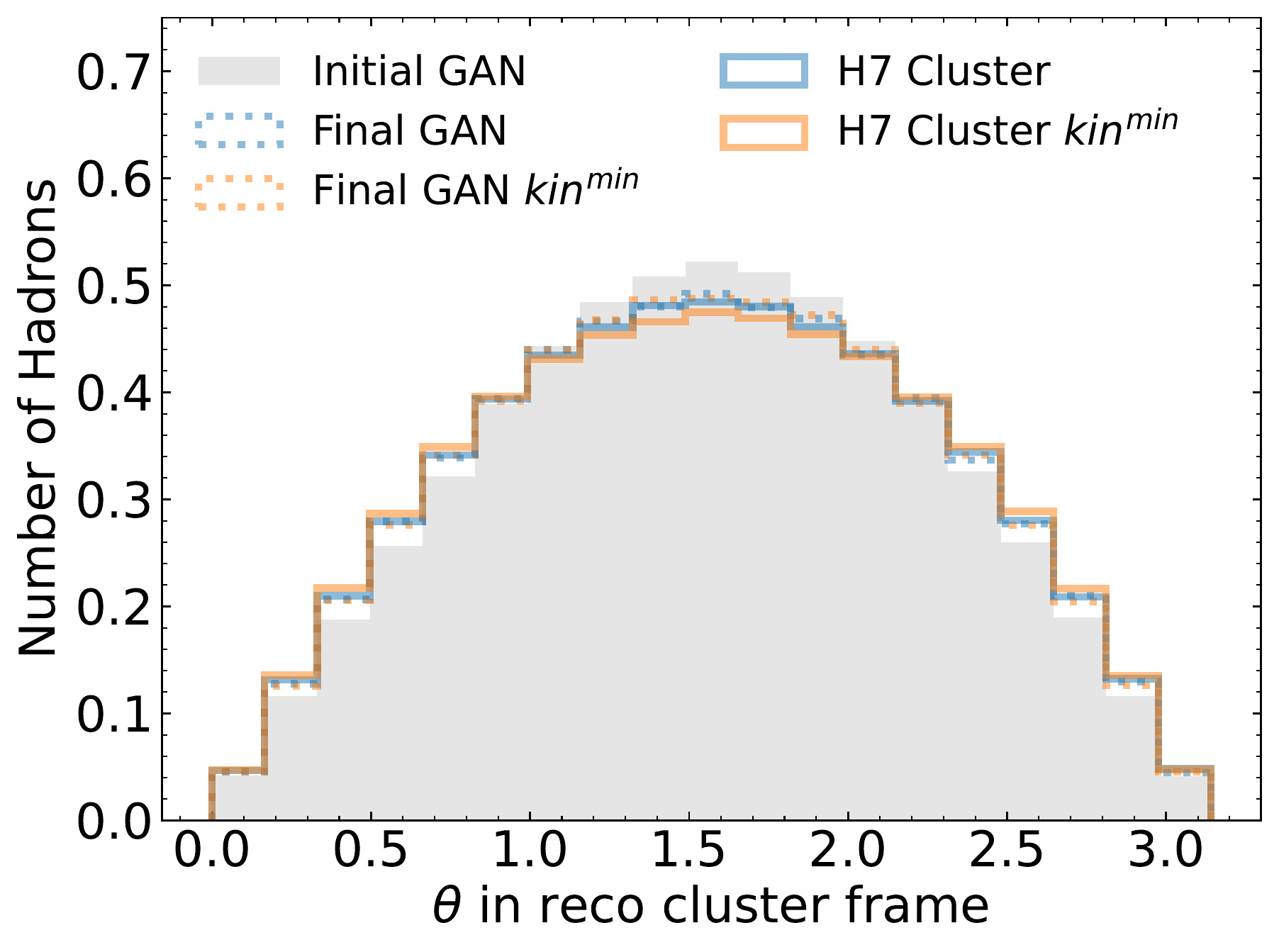}\\
    \caption{Top: A histogram of the number of hadrons.  Since each cluster decays into two pions, the number of hadrons is an even integer. Middle: $\Delta R$ between a given hadron and its nearest neighbor in $\phi-\eta$ in the lab frame (left) and the ratio of energies between a given hadron and its neighbor (right).  Bottom: The $\phi$ (left) and $\theta$ (right) of the hadrons in the reconstructed cluster frame.}
    \label{fig:fits}
\end{figure}

Figure~\ref{fig:fits} goes beyond the direct inputs and outputs by studying derived, but measureable, quantities.  The first plot in Fig.~\ref{fig:fits} is the number of hadrons.  Since we restrict our attention to $1\rightarrow 2$ decays only, the number of hadrons is an even number, with a mode of 12.  It is not possible to uniquely pair observed hadrons with their partner from the same cluster decay, but we can approximate the combination using nearest neighbor information.  In particular, since the hadron masses are small compared to the typical cluster energy in the lab frame, the two hadrons tend to be close together in phase space.  For all hadrons, we assign a hadron neighbor as the particle that minimizes\footnote{This metric is most relevant for hadron colliders, but we use it here for simplicity.  Similar results hold for $\Delta\theta$ instead of $\Delta\eta$.} $\Delta R^2=\Delta\phi^2+\Delta\eta^2$.  A histogram of the resulting $\Delta R$ distribution is shown in the middle left plot of Fig.~\ref{fig:fits}.  The peak is at about 0.1, with most hadrons having a neighbor less than 0.1.  While there is some difference between models in the $\Delta R$ distribution, a most distinguishing observable is the energy sharing between hadrons in the reconstructed cluster (middle right of Fig.~\ref{fig:fits}).  The nominal Herwig has more equal sharing of energy, while the alternative Herwig sample is much more asymmetric.  The GAN models are able to match these trends, which both differ significantly from the initialized and untrained GAN model.  Future GAN models could be improved by adding in these features to the discriminator directly.

Additionally, we consider properties of the hadrons in the reconstructed cluster frame (bottom row of Fig.~\ref{fig:fits}).  Since the reconstructed clusters are not exactly the true clusters, the $\phi$ and $\theta$ distributions do not exactly match the top row of Fig.~\ref{fig:fits_inputs}, although they are qualitatively similar.  The distribution of $\phi$ is more discriminating between models, where the GAN models perform well, except near the edge of phase space where both GAN model match the nominal Herwig events.  

A key advantage of this fitting protocol over other methods is that it can accommodate unbinned and high-dimensional inputs.  It would be possible to replace our neural network discriminator (and cross-entropy loss) with a $\chi^2$ fit to binned histograms, like the ones in Fig.~\ref{fig:fits_inputs} (bottom) and~\ref{fig:fits}, which are all observable in the lab frame.  However, this would be a highly non-trivial modification to our setup and would necessarily be less effective.  Comparing with standard tools that process low-dimensional and binned inputs would likely be inconclusive because we will not know if the difference in performance is from the tool or from the less information contained in the data.  

As a compromise in order to quantify the information gained from using our discriminator setup, we use a set of auxiliary classifiers.   Our nominal setup is represented by our discriminator trained on the same inputs as our GAN model and to distinguish the two Herwig cluster model variations.  The information content is represented by the area under the Receiver Operating Characteristic (ROC) curve or AUC, which is a standard metric for information content.  An AUC of 0.5 means there is no useful information and an AUC of 1 means that the models can be exactly distinguished.  For comparison, we compute the AUC also of the single observables in Fig.~\ref{fig:fits_inputs} (bottom) and Fig.~\ref{fig:fits}.  We do not bin these observables to avoid arbitrary binning choices and assume (which is conservative) that the bins of any actual measurement would be chosen to be maximally effective for this task.  Technically, the AUC for single observables is computed by scanning over the observable to determine the true positive rate versus the false positive rate.  

Since a threshold cut may not be optimal for all observables, we have also checked how the results change if we train a simple Boosted Decision Tree (BDT) using sklearn~\cite{scikit-learn}.  We find that the BDT-based AUCs (including for the neural network as an observable) are consistent with the non-BDT ones.  Numerically, the AUCs are as follows: neural network: 0.77, energy ratio (Fig.~\ref{fig:fits} upper right): 0.55, $\Delta R$ (Fig.~\ref{fig:fits} upper left): 0.53, rest frame $\theta$ (Fig.~\ref{fig:fits} lower right): 0.51, rest frame $\phi$ (Fig.~\ref{fig:fits} lower left): 0.51, $p_x$ (Fig.~\ref{fig:fits_inputs} lower right): 0.54, $E$ (Fig.~\ref{fig:fits_inputs} lower left): 0.57.  The information content accessible to the neural network far exceeds the information in any of the individual observables.

\clearpage

\section{Conclusions and Outlook}
\label{sec:conclusion}

We have presented a setup for fitting deep generative hadronization models to data.  The main challenge we have addressed is the lack of truth labels connecting partons and hadrons, which were used by previous deep generative hadronization models~\cite{Ilten:2022jfm,Ghosh:2022zdz}.  In order to address this challenge, we used a two-level Generative Adversarial Network (GAN) setup, where the generator acts at parton level and the discriminator acts on hadron level.  Since there is no natural order to the hadrons, the discriminator is a classifier based on the Deep Sets architecture that can process variable-length and permutation-invariant inputs.  We have shown that we can fit this model to two variations of the Herwig cluster hadronization model.  The GAN is able to reproduce Herwig well, with additional refinement and optimization required in the future to improve the prevision further.

While this represents a significant step towards realizing a deep generative hadronization model, there are still other aspects to address.  We have restricted our attention to pions, but a complete model will need to generate the full spectrum of hadrons in addition to kinematic information.  Additionally, we have started from clusters decaying to two hadrons, while in reality, more complex arrangements are possible.  In fact, we ran a test to fit the string model in Pythia using our setup\footnote{For this, we used exactly the same partons as in the Herwig dataset and ran the string model in Pythia, modified to only produce pions.}, but the cluster model is not flexible enough.  Modifications that allow for more general parton to hadron mappings, including variable-length generation~\cite{Kansal:2021cqp,Buhmann:2023pmh,Kach:2022qnf,Verheyen:2022tov,Leigh:2023toe,Mikuni:2023dvk,Finke:2023veq,Buhmann:2023bwk}, will be required in the future. In particular, we would not take pre-confinement as a starting point and instead also model the combination of partons with a neural network (so partons to hadrons instead of clusters to hadrons).  Such a model would have the capacity to mimic the cluster or string models as well as go beyond either model.  Such an architecture could be swapped out for our generator and use our same GAN setup to do the final fit.

Once we have a full model, there is a question of which data to use for the fit.  Traditionally, hadronization models have been fit to histograms (binned differential cross section measurements) from $e^+e^-$ data using tools like Professor~\cite{Buckley:2009bj} and other automated tuning protocols~\cite{Ilten:2016csi,Andreassen:2019nnm,Wang:2021gdl}.  However, these approaches may need to be modified since the parameter space of the models is much bigger.  One possibilitiy is to use a variation of Unbinned Profiled Unfolding (UPU)~\cite{Chan:2023tbf}, which uses histograms to steer neural networks with a two-level fit for unfolding.  The reweighting function in UPU could be replaced with the hadronization model.  Another possibility is to start with unbinned data, as is now possible with machine learning-based unfolding methods~\cite{Arratia:2021otl,Datta:2018mwd,Andreassen:2019cjw,Andreassen:2021zzk,bunse2018unification,Ruhe2019MiningFS,Bellagente:2019uyp,Howard:2021pos,Bellagente:2020piv,Vandegar:2020yvw,Backes:2022vmn}.  There are also now first unbinned cross section measurements~\cite{H1:2021wkz,LHCb:2022rky,H1prelim-22-034,H1prelim-22-031,H1:2023fzk}, although none are currently published without binning~\cite{Arratia:2021otl}.  There are not yet any unbinned measurements from $e^+e^-$, but results from deep inelastic scattering may be effective, since they share many of the features of $e^+e^-$ that makes them particularly clean with respect to hadron colliders.

While there are still multiple components needed to arrive at a complete ML-based hadronization model, the program ahead is well-motivated.  Current models are excellent, but the additional flexibility of neural networks will allow us to improve the precision on hadronization modeling so for precise measurements that are affected by these uncertainties.  With improvements in machine learning models, it may also be possible to use these tools to learn more about hadronization itself, which remains a key research topic in nuclear physics.

\section*{Software and Datasets}

\noindent The code for this paper can be found at \href{https://github.com/hep-lbdl/hadml/releases/tag/1.0.0}{https://github.com/hep-lbdl/hadml/releases/tag/1.0.0} \cite{jay_chan_2023_7964342}. The data sets are hosted on Zenodo at Ref.~\cite{jay_chan_2023_7958362}.

\section*{Acknowledgments}
We thank Aishik Ghosh for many useful discussions. The work of AS is funded by grant no. 2019/34/E/ST2/00457 of the National Science Centre,
Poland and the Priority Research Area Digiworld under the program Excellence Initiative
– Research University at the Jagiellonian University in Cracow.
JC, BN and XJ are supported by the U.S.\ Department of Energy (DOE), Office of Science under contract number DE-AC02-05CH11231. JC is supported by the
DOE, Office of Science under contract DE-SC0017647.




\FloatBarrier
\bibliographystyle{JHEP}
 \bibliography{HEPML,other-refs}

\end{document}